# QoS and Resource aware Security Orchestration System

Miloud Bagaa\*, Tarik Taleb\*†, Jorge Bernal Bernabe‡, Antonio Skarmeta‡
\* Aalto University, Finland, {miloud.bagaa, tarik.taleb}@aalto.fi
†University of Oulu, 90570 Oulu, Finland.
‡ Department of Communications and Information Engineering, University of Murcia,
Spain {jorgebernal, skarmeta}@um.es

*Abstract*—Network Function Virtualization (NFV) and Software Distributed Networking (SDN) technologies play a crucial role in enabling 5G system and beyond. A synergy between these both technologies has been identified for enabling a new concept dubbed service function chains (SFC) that aims to reduce both the capital expenditures (CAPEX) and operating expenses (OPEX). The SFC paradigm considers different constraints and key performance indicators (KPIs), that includes QoS and different resources, for enabling network slice services. However, the large-scale, complexity and security issues brought by these technologies create an extra overhead for ensuring secure network slicing. To cope with these challenges, this paper proposes a cost-efficient optimized SFC management system that enables the creation of SFCs for enabling efficient and secure network slices. The proposed system considers the network and computational resources and current network security levels to ensure trusted deployments. The simulation results demonstrated the efficiency of the proposed solution for achieving its designed objectives. The proposed solution efficiently manages the SFCs by optimizing deployment costs and reducing overall end-to-end delay

## I. INTRODUCTION

The recent advances in network softwarization have staggeringly benefited the 5G system and beyond by increasing the network elasticity and flexibility and reducing capital expenditures (CAPEX) operating expenses (OPEX). Network softwarization leverages both network function virtualization (NFV) and software-defined networking (SDN) paradigms to enable the utilization of common-off-the-shelf (COTS) hardware. The latter is characterized by its fast software upgrade rather than costly hardware replacement and hence enlarges different use-cases by exploiting the same device. NFV leverages virtualization techniques to deploy virtual network functions (VNFs) (e.g., router, switches, firewalls, proxies, AAA services, Virtual-HoneyNet) that run on top of general-purpose hardware. Meanwhile, SDN technology dynamically and effectively connects these VNFs by setting-up service function chains (SFCs) belonging to different service providers (SPs).

The fact of sharing the physical resources across SPs makes it more complex to orchestrate, schedule, and chaining the VNFs. The management of SFC is heightened when dealing with SDN/NFV-based IoT networks, where different virtualized and software clouds and edges are involved. The heterogeneity of these players imposes challenges to manage and optimize the overall end-to-end delay as different heterogeneous virtual network infrastructures in the SFC are required.

To deal with the above challenges, in this paper, we suggest a cost-efficient SFC deployment and management framework that aims to ensure the desired KPIs that include the QoS and security while reducing the overall cost. The proposed mechanism considers the implementation and management of SFCs that are potentially deployed across IoT domains, inter-clouds, and cross-network segments. Therefore, ensuring efficient management of SFC in a virtualized and software IoT network is a challenging process that is targetted by this paper. To this aim, unlike the current state of the art, besides the QoS and actual capacities of VNFs in terms of resources (CPU, RAM, and storage), the proposed solution also considers the network security levels (enforced security channels mechanisms across the VNFs in the SFC). The algorithm finds a predefined blueprint configuration for the SFC, and targeted KPIs, and optimizes the targeted KPIs in terms of link security level, QoS, delay, and bandwith.

The remaining of the paper is organized as follows. Section II summarizes the related works. Section III introduces the main idea and the problem formulation targeted by this paper. The optimization algorithm is described in section IV, while section V shows the simulation results and its discussion. Finally, section VI concludes the article.

## II. RELATED WORK

The security is one of the main challenges facing the management and orchestration of SFC [1], [2], and VNFs. The deployment and management of SFC in a dynamic environment is a highly challenging process that attracted the attention of industry and academia in the past. Authors in [3] have suggested a basic formulation of VNF scheduling that considers the processing delay at different VNFs in the chain. Meanwhile, authors in [4] bring up simple algorithms that aim at minimizing the time, the cost, and the revenue in function mapping and scheduling.

Authors in [5] have studied the problem of VNFs allocation and deployment in a distributed environment. However, the authors do not consider the placement of these VNFs in different SFCs. Similarly, authors in [6] have addressed the service chaining optimization challenge. However, they do not find the end-to-end delay when deploying the SFCs. Similarly, in [7], authors formalize the VNF placement and SFC problem as an Integer Linear Programming (ILP). They solve the optimization problem by considering the cost and the distance across clients and VNFs. However, the proposed optimization does not find other resources and security aspects, as addressed in our model. Moreover, in contrast to this solution, our solution considers cost-efficient life cycle management when allocating computational and network resources. Some other related research papers mainly focus on different kinds of QoS metrics for improving performances, such as bandwidth allocation optimization for VNFs usage [8]. Meanwhile, authors [9] authors analyze the VNF scheduling and resource allocation

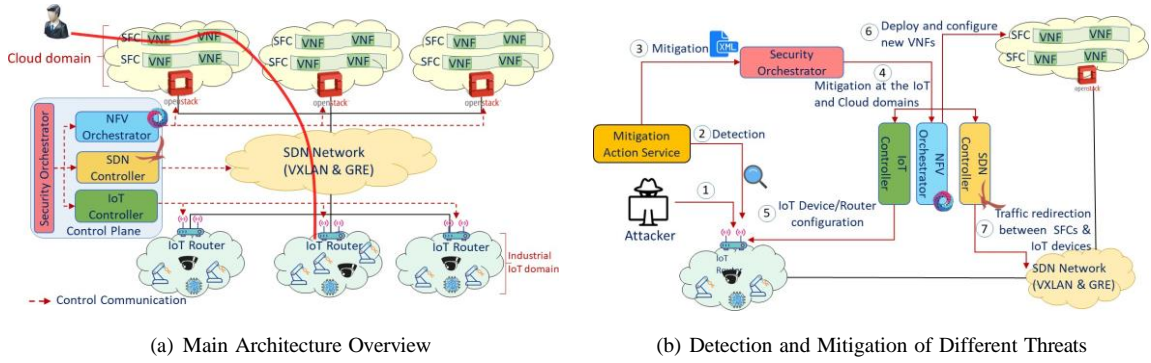

(a) Main Architecture Overview

(b) Detection and Mitigation of Different Threats

Fig. 1: Envisioned System: Architecture and Security Enablers

with SFCs and provide resource optimization solutions. They define the scheduling decisions and chaining based on service transmission and VNF's processing delays. However, unlike the proposed algorithm, they do not consider security aspects in the formulation.

Authors in [10] have suggested a reliability-aware SFC solution that aims to provision VNFs with delay guarantees. To solve the problem, they have suggested the use of a mixed-integer linear program (MILP) for VNFs allocation that maximizes reliability and end-to-end delays. From another side, Artificial Intelligence has been leveraged in [11] for ensuring efficient VNF scheduling. Unfortunately, the work considers only the processing and transmission delay when scheduling the SFCs. Meanwhile, Long Qu et al. [12] have proposed a model for SFC orchestration in 5G networks. The solution has been modeled as ILP, and then heuristic algorithms have been suggested. However, the proposed solutions consider only the minimization of the path length and cost of computing resources.

## III. NETWORK MODEL AND PROBLEM FORMULATION

### A. Main idea

Fig. 1 depicts the envisioned system targeted by this paper. As illustrated in Fig. 1, the user can access to IoT domains via SFCs that are deployed on multiple administrative and technological domains. The management of these SFCs happens at the security orchestrator (SO). The latter can manage the connection and mitigate the attacks by leveraging: $i$) NFV orchestrator; $ii$) SDN controller; $iii$) IoT controller. In the the suggested framework, when an attack is detected by the mitigation action service (MAS) component, a mitigation request (security policy) should be generated and sent to the security orchestrator (SO) (Fig. 1(b), Steps 1⊢ 3). The latter should run an optimization algorithm, suggested in this paper, to mitigate the attacks by deploying or redirecting traffic through existing SFCs by taking into account the actual resources, the requested QoS, and security aspects.

Furthermore, the new SFCs can use either complete new security VNF instances (VNFIs) or already existing ones. When designing and deploying new SFCs, the SO considers the used security VNFs and their communication links for ensuring the desired security and QoS. For instance, the communication between different clouds, VNFs, users, and devices, is characterized by unusual delay, bandwidth, jitters, and security levels. For instance, the communication link within the same cloud has a high-security level, while the one between users, devices, and inter clouds would have more vulnerability. Moreover, the link security levels can be affected by using security mechanisms, such as IPsec, SSL or even DTLs when targeting IoT environments. Later, the SO will mitigate the attacks by communicating with IoT controllers, NFV orchestrators, and/or SDN controllers (Fig. 1(b), Step 4).

The attacks would be mitigated at the edge by communicating with the IoT controller. For instance, the malicious IoT devices could be reconfigured (using protocols such as CoAP or LWM2M), shut-off, rebooted and even flushed, thereby preventing the attacks to be spread from a domain to another (Fig. 1(b), Step 5). The attacks could also be mitigated via the SDN controller by either preventing the attacks, e.g. adding new flow rules to drop the traffic from the attacker or limiting the bandwidth (Fig. 1(b), Step 6). Finally, the attacks can also be mitigated by deploying or redirecting traffic through SFCs (Fig. 1(b), Steps 6 ⊢ 7).

TABLE I: Summary of Notations.

| Notation | Description |
|---|---|
| U | The set of users in the network. |
| D | The IoT devices deployed in a distributed IoT infrastructure. |
| $V_C$ | The set of clouds or edges existing in the network. |
| R | The set of all the resources. R could equal to {RAM, CPU, DISK}. |
| $\Delta_r(c)$ | A parameter that shows the amount of resource $r \in R$ available at the cloud/edge $c \in V_C$. |
| $\Theta$ | The set of service function chaining (SFCs) that serves the users. |
| $\Phi_\theta$ | The users of the SFC $\theta \in \Theta$. |
| $\lambda_\theta$ | The expected traffic that will be exchanged between the users $\Phi_\theta$ and the IoT devices D. |
| $Y_\theta$ | The list of VNFs forming the SFC $\theta \in \Theta$. |
| $\Pi$ | The set of VNFs' types supported in the system, such as firewall, load balancer,...etc. |
| $\pi_v$ | The type of the VNF $v \in Y_\theta$, such that $\pi_v \in \Pi$. |
| V | The VNF instances that should be instantiated to serve the VNFs $\bigcup_{\theta \in \Theta} Y_\theta$. |
| $\zeta(v)$ | All the VNFs that have the same type like the VNF $v \in Y_\theta$. $\zeta(v)$ is defined formally as follow: $\zeta(v) = \{v' : \forall v' \in \bigcup_{\theta' \in \Theta} Y_{\theta'} \wedge \pi_v = \pi_{\theta'}\}$. |
| $\Psi(v)$ | All the VNFs that have conflicts with the VNF $v \in Y_\theta$. |
| F | The list of flavors that can be used by different VNFs. Each flavor $f \in F$ specifies the amount of all resources $r \in R$ that should be used by a VNF instance. |
| $\delta_r(f)$ | The amount of resources $r \in R$ used by the flavor $f \in F$. |
| $\delta_p(f)$ | The price of using the flavor $f \in F$. |
| $\Omega(u)$ | The set of VNFs that would be hosted at the VNFI $u \in V$. |
| $\Gamma(.)$ | $\Gamma(\pi_u, f, \sum_{\theta \in \Theta, v \in Y_\theta \cap \Omega(u)} \lambda_\theta)$ denotes the average processing delay would be expected from the VNFI $u$ to process one packet. |

## B. Problem formulation

We denote by $U$ a set of users that interested in different services offered by IoT devices $D$ deployed in a distributed IoT infrastructure. Let $(V, E, W)$ denote the underlying network that consists of a set of IoT devices and clouds. $V$ consists of: $i$) Clouds or edges ($V_C$); $ii$) Distributed IoT infrastructure ($V_I$); $iii$) End-users ($V_U$) that represent the (radio) access nodes ((R)AN) from whereby the users access. Each cloud/edge $u \in V_C$ is characterized by limited storage and computation resources, including CPU, RAM, and DISK. Let $\Delta_r(u)$ be a vector that shows the resources of the cloud $u \in V_C$. For the sake of simplicity, $E$ represents the end-to-end connections between $V_U \cup V_I \cup V_C$ by considering an overlay network and by making an abstraction on the intermediate routers and switches (backbone). However, the suggested model is orthogonal and can be easily adapted and extended to consider the intermediate network components if needed. Meanwhile, $W$ denotes the characteristics of the links that include the bandwidth, delay, and security level. Let $W^B$, $W^L$ and $W^S$ denote the characteristics of the links $E$ in terms of bandwidth capacity, propagation delay and security level, respectively.

We denote by $\Theta$ the set of SFCs in the network, such that each SFC $\theta \in \Theta$ serves a set of users. Let $\Phi_\theta$ denote the set of users of the SFC $\theta \in \Theta$. The SO would have the information about the expected traffic that would be exchanged between the users $\Phi_\theta$ and the IoT devices. Let $\lambda_\theta$ denote the expected traffic that will be exchanged between the users $\Phi_\theta$ and the IoT devices.

In the proposed model, we leverage the strength of SDN technology to interconnect the users with their IoT devices. Formally, the communication between a user and its IoT devices happens through a specified SFC $\theta \in \bar{\Theta}$ that consists of a list of VNFs $Y_\theta$. We denote by $\Pi$ the set of VNFs types supported by the system. We denote by $\pi_v$ the type of the VNF $v \in Y_\theta$, such that $\pi_v \in \Pi$. We denote by $(v_i, v_j) \in Y_\theta$ a two consecutive VNFs in the SFC $\theta \in \bar{\Theta}$. In the blueprint, the KPIs of each SFC $\theta \in \bar{\Theta}$ consider the end-to-delay and bandwidth, as well as the security levels of the links used for interconnecting $\theta$'s components. While the two former metrics target the quality of experience (QoE), the latter targets the security level of $\theta$. Let $\sigma^L_\theta$, $\sigma^B_\theta$ and $\sigma^S_\theta$ denote the $\theta$'s end-to-end delay, bandwidth and security level, respectively. Let $V$ represent the instantiated VNFs in different clouds. $V$ represents the VNF instances (VNFI) that would be used by the VNFs $\Theta$. Note that more than one VNF in ($\bar{\Theta}$) could use the same VNFI in ($V$). Based on the observation that each new VNF would be maximum instantiated at one VNFI, we have $|V| \leq |\bigcup_{\theta \in \Theta} Y_\theta|$.

In the proposed model, one VNFI can be shared with multiple SFCs, and hence two VNFs $(v_1, v_2) \in Y_{\theta_1} \times Y_{\theta_2}$, that belong to two different SFCs $\theta_1, \theta_2 \in \bar{\Theta}$, could use the same VNFI $u \in V$ if and only if the following conditions hold: $i$) They have the same type $\pi_{v_1} = \pi_{v_2}$; $ii$) There are no conflicts in the configuration of the two VNFs $v_1$ and $v_2$. Let $\zeta(v)$ denote all the VNFs that have the same type like $v$. Formally $\zeta(v)$ is defined as follows: $\zeta(v) = \{v' : \forall v' \in \bigcup_{\theta' \in \Theta} Y_{\theta'} \wedge \pi_v = \pi_{v'}\}$. Meanwhile, $\Psi(v)$ denotes all the VNFs that have conflicts with the VNF $v \in Y_\theta$. The SO receives this list from the "Policy Interpreter" component.

Formally, the candidate list of VNFs that can be hosted in the same VNFI with the VNF $v \in Y_\theta$ is $\zeta(v) - \Psi(v)$. Let $R$ denote the resources that are considered. $R$ can be defined as follows: $R = \{RAM, CPU, STORAGE\}$. Also, we denote by $F$ the list of available flavors in the network. Each flavor $f \in F$ has a specified amount of resources should be used by a VNF instance. Let $\delta_r(f)$ denote the amount of resources $r \in R$ should be used by the favor $f$. We denote by $\Omega(u)$ the set of VNFs that would be hosted at the VNFI $u \in V$. Formally, there is a correlation between the resources used by a VNFI $u \in V$, the amount of traffic treated by that VNFI, and the expected services offered by that VNFI in terms of computation and QoS. While the amount of traffic negatively affects the QoS, the amount of resources has a positive impact on the QoS [13], [14].

Let $\pi_u$ denote the VNF instance type. Note that a VNFI $u$ whose type is similar to the VNFs that host. Let $\Gamma(\pi_u, f, \sum_{\theta \in \Theta, v \in Y_\theta \cap \Omega(v)} \lambda_\theta)$ denote a function that returns the average expected processing delay for handling the data traffic $\sum_{\theta \in \Theta, v \in Y_\theta \cap \Omega(v)} \lambda_\theta$ by a VNFI $u$ that uses the flavor $f \in F$ and has a type $\pi_u \in \Pi$. For the sake of simplicity, we consider that $\Gamma(.)$ is a linear function in respect to the amount of data $\lambda_\theta$.

## IV. PROPOSED SOLUTION

In this section, we present the solution that optimizes the deployment of SFCs by taking into account the QoS and the security. In this section, we first start by defining the different variables, and then we will describe the various constraints.

### A. VNF and VNFI relationship constraints

We define by $X_{v,u}$ a decision Boolean variable that shows if the VNF $v \in Y_\theta$ of the SFC $\theta \in \Theta$ will use the VNFI $u \in V$. Formally, $X_{v,u} = 1$ if $v$ uses VNFI $u$, otherwise $X_{v,u} = 0$. $X_{v,u}$ should be subject to the following constraint. VNFs in each SFC $\theta$ should use only one VNFI. Each VNF, $v \in Y_\theta$ for $\theta \in \hat{\Theta}$, should use only and only one VNFI $u$ in the network.

$$\forall \theta \in \Theta, \forall v \in Y_\theta : \sum_{u \in V} X_{v,u} = 1 \quad (1)$$

From another side, let $B_{\theta,u}$ denote a Boolean decision variable that shows if an SFC $\theta \in \Theta$ would use VNFI $u \in V$. Formally, $B_{\theta,u} = 1$ if $u$ is used by $\theta$, otherwise $B_{\theta,u} = 0$. For preventing the loop in the SFC $\theta$, VNFs ($\forall v \in Y_\theta$) of that SFC should use the VNFI $u$ maximum once as mentioned by constraint (2).

$$\forall u \in V, \forall \theta \in \Theta : \sum_{v \in Y_\theta} X_{v,u} = B_{\theta,u} \quad (2)$$

The following constraints ensure that a VNF $v$ should not share a VNFI $u \in V$ with another VNF with whom has a conflict or does not have the same type.

We define by $Y_{v,c}$ a decision Boolean variable that shows if a VNF $v \in Y_\theta$ for $\theta \in \Theta$ is hosted at the cloud $c \in V_C$ or not. $Y_{v,c}$ equals to 1 if the VNF $v$ is hosted in the cloud $c$, otherwise $Y_{v,c} = 0$. Formally a VNF $v$ is hosted at a cloud $c \in V_C$ if its VNFI is running on top of that cloud. Each VNF should run on top only one cloud at a given time. Constraint (3) ensures that each VNF should be deployed only on one cloud.

$$\forall \theta \in \Theta, \forall v \in Y_\theta : \sum_{c \in V_C} Y_{v,c} = 1 \quad (3)$$

We define also by $A_{u,\pi}$ a decision Boolean variable that shows if the VNFI $u \in \hat{V}$ has the type $\pi \in \Pi$. Formally, $A_{u,\pi} = 1$ if and only if VNFI $u$ has the type $\pi$. Each VNF $v \in Y_\theta$ for $\theta \in \Theta$ that uses VNFI $u \in V$ should use the same VNF type as mentioned in constraint 4.

$$\forall \theta \in \Theta, \forall v \in Y_\theta, \forall u \in V, \forall \pi \in \Pi:$$
$$\pi_v \times X_{v,u} = \pi \times A_{u,\pi} \times X_{v,u} \quad (4)$$

Unfortunately, the constraint (4) is not linear due to the part $A_{u,\pi} \times X_{v,u}$. In order to make this constraint linear, we replace the constraint (4) with the constraints (5) and (6).

$$\forall \theta \in \Theta, \forall v \in Y_\theta, \forall u \in V, \forall \pi \in \Pi:$$
$$\pi_v \leq \pi + (2 - X_{v,u} - A_{u,\pi}) \times M \quad (5)$$

, where M is a big number ($M \approx \infty$).

$$\forall \theta \in \Theta, \forall v \in Y_\theta, \forall u \in V, \forall \pi \in \Pi:$$
$$\pi \leq \pi_v + (2 - X_{v,u} - A_{u,\pi}) \times M \quad (6)$$

The following constraint ensures that each VNFI $u \in V$ should have only one type if and only of it is deployed, otherwise it should not have any type.

$$\forall u \in V: \sum_{\pi \in \Pi} A_{u,\pi} = \sum_{c \in V_C} U_{u,c} \quad (7)$$

Constraints (5), (6) and (7) ensure VNFs should have the same type of their VNFIs, which prevent the deployment of the VNFs that have different types on the same VNFI.

### B. VNFI and cloud relationship constraints

We also define by $U_{u,c}$ a Boolean variable that shows if the VNFI $u \in V$ would be deployed in the cloud/edge $c \in V_C$. Formally, $U_{u,c} = 1$ if and only if $u$ will be deployed in the cloud $c$, otherwise $U_{u,c} = 0$. First of all, a VNFI $u \in V$ should be deployed if and only if it is used at least by one VNF as presented by the constraint (8).

$$\forall u \in V, \forall \theta \in \Theta, \forall v \in Y_\theta: \sum_{c \in V_C} U_{u,c} \geq X_{v,u} \quad (8)$$

Also, a VNFI $u \in V$ should be not deployed if it is not used by any VNF in the network as shown in the constraint (9).

$$\forall u \in V: \sum_{c \in V_C} U_{u,c} \leq \sum_{\theta \in \Theta, v \in Y_\theta} X_{v,u} \quad (9)$$

Finally, each VNFI $u \in V$ should be deployed at most at one cloud as shown in the constraint (10).

$$\forall u \in V: \sum_{c \in V_C} U_{u,c} \leq 1 \quad (10)$$

A VNF should use the same cloud where its VNFI is deployed. Constraint (11) ensures that each VNF should use the same cloud as its VNFI. While constraint (12) ensures that a VNF $v$ should not use a cloud $c$ if its VNFI is not running on that cloud.

$$\forall c \in V_C, \forall u \in V, \forall \theta \in \Theta, \forall v \in Y_\theta:$$
$$Y_{v,c} \geq X_{v,u} \times U_{u,c} \quad (11)$$

$$\forall c \in V_C, \forall u \in V, \forall \theta \in \Theta, \forall v \in Y_\theta:$$
$$Y_{v,c} \leq \sum_{u \in V} X_{v,u} \times U_{u,c} \quad (12)$$

Unfortunately, the constraint (11) and (12) are not linear. To make the optimization linear, we have replaced the constraints (11) and (12) with the following variables and constraints. First we define the variable $Y_{v,u,c}$ for $v \in Y_\theta$, $u \in V$, and $c \in V_C$, respectively. Formally, $Y_{v,u,c} = 1$ if and only if VNF $v$ uses VNFI $u$ that is deployed in the cloud/edge $c$. Otherwise, $Y_{v,u,c} = 0$. Then, we replace the constraint (11) and (12) with the following constraints:

Constraints (13) and (14) ensure that $Y_{v,u,c} = 0$ if $v$ does not use $u$ (i.e., $X_{v,u} = 0$) or $u$ is not deployed at cloud $c$ (i.e., $U_{u,c} = 0$).

$$\forall c \in V_C, \forall u \in V, \forall \theta \in \Theta, \forall v \in Y_\theta:$$
$$Y_{v,u,c} \leq U_{u,c} \quad (13)$$

$$\forall c \in V_C, \forall u \in V, \forall \theta \in \Theta, \forall v \in Y_\theta:$$
$$Y_{v,u,c} \leq X_{v,u} \quad (14)$$

Constraint (15) ensures that $Y_{v,u,c} = 1$ if and only if both variables $X_{v,u}$ and $U_{u,c}$ equal to 1.

$$\forall c \in V_C, \forall u \in V, \forall \theta \in \Theta, \forall v \in Y_\theta:$$
$$Y_{v,u,c} \geq X_{v,u} + U_{u,c} - 1 \quad (15)$$

Meanwhile, constraint (16) ensures that the VNF $v$ is deployed in cloud $c$ (i.e., $Y_{v,c} = 1$) if $Y_{v,u,c} = 1$.

$$\forall c \in V_C, \forall u \in V, \forall \theta \in \Theta, \forall v \in Y_\theta:$$
$$Y_{v,c} \geq Y_{v,u,c} \quad (16)$$

Constraint (17) ensures that a VNF $v$ is not deployed in a cloud $c$ (i.e., $Y_{v,c} = 0$) if it does not use any VNFI deployed at cloud $c$.

$$\forall c \in V_C, \forall \theta \in \Theta, \forall v \in Y_\theta:$$
$$Y_{v,c} \leq \sum_{u \in V} Y_{v,u,c} \quad (17)$$

Meanwhile, the constraint (18) ensures that each VNF $v$ should be hosted at a cloud.

$$\forall \theta \in \Theta, \forall v \in Y_\theta: \sum_{c \in V_C} Y_{v,c} = 1 \quad (18)$$

### C. Resource aware constraints

We define by $\varphi_{u,f}$ a decision Boolean variable that shows if VNFI $u \in V$ would use a flavor $f \in F$. Formally, $\varphi_{u,f} = 1$ if $u$ uses flavor $f$, otherwise $\varphi_{u,f} = 0$. First of all, each VNFI should use only one flavor if it is deployed as shown in the constraint (19).

$$\forall u \in V: \sum_{f \in F} \varphi_{u,f} = \sum_{c \in V_C} U_{u,c} \quad (19)$$

The constraint (20) ensures that each cloud/edge should not overloaded. The amount of resource $r \in R$ should not exceed the capacity $\Delta_r(c)$ of the cloud $c \in V_C$ is not overloaded.

$$\forall r \in R, \forall c \in V_C: \sum_{f \in F} \sum_{u \in V} \delta_r(f) \times \varphi_{u,f} \times U_{u,c} \leq \Delta_r(c) \quad (20)$$

Unfortunately, the constraint (20) is not linear due to the part $\varphi_{u,f} \times Y_{u,c}$. To convert the optimization to linear

optimization. we replace the constraint (20) by the following constraints and variables:

First, we define the variable $C_{u,c,f}$ that shows if the VNFI $u$ uses the flavor $f$ in the cloud $c$. Formally, $C_{u,c,f} = 1$ if and only if $u$ uses flavor $f$ in cloud $c$. Also, we replace the constraint 20 with the following constraints.

$$\forall c \in V_C, \forall f \in \mathcal{F}, u \in U : C_{u,c,f} \leq C_{u,c} \quad (21)$$

$$\forall c \in V_C, \forall f \in \mathcal{F}, u \in U : C_{u,c,f} \leq \varphi_{u,f} \quad (22)$$

$$\forall c \in V_C, \forall f \in \mathcal{F}, u \in U : C_{u,c,f} \geq C_{u,c} + \varphi_{u,f} - 1 \quad (23)$$

$$\forall r \in Y, \forall c \in V_C : \sum_{f \in F} \sum_{u \in V} \delta_r(f) \times C_{u,c,f} \leq \Delta_r(c) \quad (24)$$

### D. QoS aware constraints

The following constraint ensures that the end-to-end delay in the SFCs is respected. In our model, we consider that the propagation delay dominates the computation one, and hence we ignore the processing delay. Let we define by $f^C_{c_1,c_2}$ the expected propagation delay between two clouds $c_1, c_2 \in V_C, c_1 \neq c_2$ after binding and deploying the SFCs.

$$\forall c_1, c_2 \in V_C, c_1 \neq c_2 : f^C_{c_1,c_2} = \frac{1}{y_{c_1,c_2}} \sum_{\theta \in \Theta} \sum_{(v_1,v_2) \in Y_\theta} \lambda_\theta \times Y_{v_1,c_1} \times Y_{v_2,c_2} \quad (25)$$

However, the constraint (25) is not linear due to the part $Y_{v_1,c_1} \times Y_{v_2,c_2}$. In order to make the optimization linear, we replace the constraint (25) by the following variables and constraints. First of all, we add $Y_{v_1,c_1,v_2,c_2}$ variable that equals to 1 if and only if $v_1$ and $v_2$ are hosted at $c_1$ and $c_2$, respectively. Otherwise, $Y_{v_1,c_1,v_2,c_2} = 0$.

$$\forall \theta \in \Theta, \forall (v_1, v_2) \in Y_\theta, \forall c_1, c_2 \in V_C, c_1 \neq c_2 : Y_{v_1,c_1,v_2,c_2} \leq Y_{v_1,c_1} \quad (26)$$

$$\forall \theta \in \Theta, \forall (v_1, v_2) \in Y_\theta, \forall c_1, c_2 \in V_C, c_1 \neq c_2 : Y_{v_1,c_1,v_2,c_2} \leq Y_{v_2,c_2} \quad (27)$$

$$\forall \theta \in \Theta, \forall (v_1, v_2) \in Y_\theta, \forall c_1, c_2 \in V_C, c_1 \neq c_2 : Y_{v_1,c_1,v_2,c_2} \geq Y_{v_1,c_1} + Y_{v_2,c_2} - 1 \quad (28)$$

$$\forall c_1, c_2 \in V_C, c_1 \neq c_2 : f^C_{c_1,c_2} = \frac{1}{y_{c_1,c_2}} \sum_{\theta \in \Theta} \sum_{(v_1,v_2) \in Y_\theta} \lambda_\theta \times Y_{v_1,c_1,v_2,c_2} \quad (29)$$

Then, we will define the propagation delay between two consecutive VNFs in an SFC. Let $f^V_{v_1,v_2}$ denote the propagation delay between two consecutive VNFs $(v_1, v_2) \in Y_\theta$ for $\theta \in \Theta$. Let we define $f^V_{v_1,v_2}$ for SFCs $\theta \in \hat{\Theta}$.

$$\forall c_1, c_2 \in V_C, c_1 \neq c_2, \forall \theta \in \Theta, \forall (v_1, v_2) \in Y_\theta : f^V_{v_1,v_2} \leq f^C_{c_1,c_2} + (1 - Y_{v_1,c_1,v_2,c_2}) \times \mathbf{n} \quad (30)$$

$$\forall c_1, c_2 \in V_C, c_1 \neq c_2, \forall \theta \in \Theta, \forall (v_1, v_2) \in Y_\theta : f^C_{c_1,c_2} \leq f^V_{v_1,v_2} + (1 - Y_{v_1,c_1,v_2,c_2}) \times \mathbf{n} \quad (31)$$

In what follow, we will define the constraints that ensure the end-to-end delay of each SFC is respected. Thus, the end-to-end delay of each SFC $\theta \in \bar{\Theta}$ does not exceed its threshold $\xi^L_\theta$.

$$\forall \theta \in \Theta : \sum_{(v_1,v_2) \in Y_\theta} f^V_{v_1,v_2} \leq \xi^L_\theta \quad (32)$$

Finally, we need to ensure that all the paths selected by the VNFs of a given SFC respect the constraints of security. Indeed, all the links that interconnect VNFs of the same SFC $\theta \in \Theta$ should have a security level higher than the security level $\xi^S_\theta$.

$$\forall c_1, c_2 \in V_C, c_1 \neq c_2, \forall \theta \in \Theta, \forall (v_1, v_2) \in Y_\theta : \xi^X_\theta \times Y_{v_1,c_1,v_2,c_2} \leq Y^S_{c_1,c_2} \quad (33)$$

### E. Final Optimization problem

In what follow, we present the global optimization solution that aims to reduce the cost while taking into account the QoS during the deployment of different SFCs.

$$\min \sum_{u \in \hat{V}} \sum_{f \in F} \delta_p(f) \times \varphi_{u,f} \quad (34)$$

S.t,

<u>VNF and VNFI relationship constraints:</u>

(1), (2), (3), (5), (6) and (7).

<u>VNFI and cloud relationship constraints:</u>

(8), (9), (10), (13), (14), (15), (16), (17) and (18).

<u>Resource aware constraints:</u>

(19), (21), (22), (23) and (24)

<u>QoS aware constraints:</u>

(26), (27), (28), (29), (30), (31), (32) and (33).

## V. PERFORMANCE EVALUATION

The proposed solution has been implemented and evaluated using Python language and Gurobi Optimizer software. In all the simulations, we have considered 15 security levels, where we have uniformly distributed them among the connection links that interconnect clouds, edges, and IoT domains. We have evaluated the suggested solution in different scenarios by running 35 repetitions. The plotted results present the mean and 95% confidence interval. First, the average cost that is needed for deploying VNFIs in public clouds for hosting the VNFs of different SFCs. Second, the average end-to-end delay perceived by different SFCs in different epochs.

The suggested solution is evaluated by varying the number of edges and the number of deployed SFCs. In the first scenario, we have varied the number of edges/clouds while fixing the number of SFCs by 4. Meanwhile, in the second scenario, we have modified the number of SFCs while fixing the number of clouds/edges by 40.

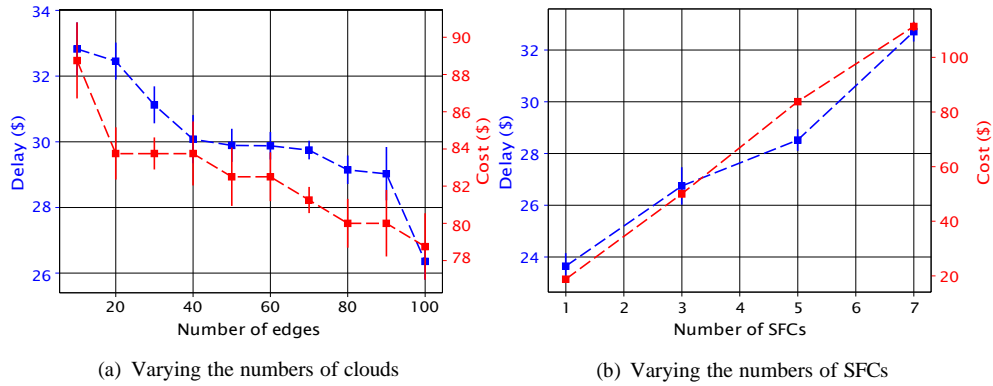

(a) Varying the numbers of clouds  (b) Varying the numbers of SFCs

Fig. 2: Cost and Average Delay

Fig. 2(a) depicts the impact of the number of edges on the cost and delay. While the delay is shown with a blue curve with the y-axis in the left, the cost is depicted with a red color curve and y-axis in the right. The first observation that we can draw from this figure is that the number of edges has a positive impact on the cost. Increasing the number of edges/clouds gives more flexibility to the solution for finding a better position for deploying the VNFIs that serve more VNFs, and hence helping more SFCs. This leads to reducing the number of VNFIs that should be used, and therefore reduce the cost. Moreover, increasing the number of edges/clouds also has a positive impact on the delay. Increasing the number of edges/clouds leads to increasing the likelihood of finding better locations for deploying different VNFIs that could offer better end-to-end delay for different SFCs.

Meanwhile, Fig. 2(b) shows the impact of the number of SFCs on the cost and delay, respectively. Increasing the number of SFCs that should be deployed leads to an increase in the number of deployed VNFs, and hence it harms the cost. We also observed that the rise in the number of SFCs hurts the end-to-end delay. This can be explained as follow: Increasing the number of SFCs leads to an increase in the probability of getting VNFs that have more resources' requirements. The resources demanding of these VNFs will limit their options to be deployed in clouds/edges that offer better delays. Also, increasing the number of SFCs leads to raising the likelihood to get SFCs with sparsely distributed users, which hurts the end-to-end delay.

## VI. Conclusion

This paper has proposed an efficient Algorithm for deploying different SFC while considering the QoS, actual capacities of VNFs in terms of resources (CPU, RAM, and storage) and current network security levels. The Algorithm has been evaluated via simulation. The obtained results demonstrate the effectiveness of the Algorithm for achieving the desired objectives. As future work we envisage to extend the Algorithm to consider SFC management also in volatile and mobile edge nodes that imposes additional physical constrains to the optimization.

## Acknowledgment

This work was partially supported by the European research project H2020 INSPIRE-5Gplus GA 871808. It has been also partially funded by AXA Postdoctoral Scholarship awarded by the AXA Research Fund (Cyber-SecIoT project). This work was partially supported by the Academy of Finland 6Genesis project under Grant No. 318927, and by the Academy of Finland CSN project under Grant No. 311654.